\documentclass{iaus}
\usepackage{graphicx}

\title[Wide binaries in the solar neighborhood] 
{Catalogs of Wide Binaries:\\ Impact on Galactic
Astronomy
}

\author[Chanam\'e]   
{Julio Chanam\'e}

\affiliation{Space Telescope Science Institute}

\pubyear{2006}
\volume{240}  
\date{9/18/06 and in revised form ??}
\jname{Binary Stars as Critical Tools \& Tests in Astrophysics}
\editors{Bill Hartkopf, Ed Guinan \& Petr Harmanec, eds.}

\newcommand\arcsec{\mbox{$^{\prime\prime\,}$}}%
\newcommand{\msun}{\,M$_{\odot\,\,}$}

\begin{document}

\def\kms{{\rm km}\,{\rm s}^{-1}}
\def\masyr{{\,\rm mas}\,{\rm yr}^{-1}}

\maketitle

\begin{abstract}

Wide binaries, particularly in large numbers and as free from
selection biases as possible, constitute a largely overlooked tool for
studying the Galaxy. The goal of this review is to highlight the
potential inherent to large samples of field wide binaries for
research on problems as varied as star formation in the early Galaxy,
the nature of halo dark matter, the evolution of the stellar halo, new
geometric distances, metallicities, masses, and ages of field stars
and white dwarfs, and much more. Using the Revised NLTT as an
illustrative example, I review the main steps in the assembly of a
large catalog of wide binaries useful for multiple applications. The
capability of cleanly separating between the Galactic disk and halo
populations using good colors and proper motions is emphasized. The
critical role of large surveys for research on wide binaries as well
as for the better understanding of the Galaxy in general is stressed
throughout. Finally, I point out the potential for assembling new
samples of wide binaries from available proper-motion surveys, and
report on current efforts of using the SDSS towards this goal.


\end{abstract}

\firstsection 
\section{Introduction}
\label{sec:intro'}

Knowledge obtained from the study of stars in binary systems has been
fundamental for astronomy, and influences almost all of its
branches. Their impact is easily illustrated with just a few examples:
the masses of individual stars (the single most important parameter in
stellar evolution) are measured with confidence only for stars in
binary systems\footnote{Stellar mass measurements for single stars
have been achieved thanks to gravitational microlensing as well
(\cite[Drake et al. 2004]{drake04}; \cite[Jiang et
al. 2004]{guangfei}).  However, the errors in this technique are not
yet competitive with those for stars in binary systems.}; eclipsing
binaries provide a primary method for measuring distances
(\cite[Paczy\'nski 1997]{pac97}) and are playing a major role in
securing the base of the cosmological distance ladder (\cite[Ribas et
al. 2005]{rib05}; \cite[Bonanos et al. 2006]{bon06}); close binary
systems harboring white dwarfs, as likely progenitors of type-Ia
supernovae, are thought to be main players driving the chemical
evolution of galaxies and the intergalactic medium (\cite[Pagel
1997]{pag97}). And the list can go on and on.

Most of that knowledge, however, has come from the study of binary
systems in close orbits, while the population of binaries at the wide
end of the distribution of orbital separations (semimajor axes
$a\gtrsim 100$\,AU) remains poorly explored. Clearly, the main reason
behind this historical bias is the much longer orbital timescale
inherent to very wide binary systems (of the order of hundreds of
years and longer), making their unambiguous identification a task that
also requires a relatively long timescale (though, fortunately, not as
long as their complete periods!). In contrast to the case of close
binaries, where some kind of variability is typically the property
that helps uncover them, a reasonably large number of very wide
binaries can not be identified via the typical photometric or
spectroscopic campaign lasting from several months to a few
years. Instead, wide binaries require accurate astrometry over
timescales of the order of decades.

Nevertheless, wide binaries hold great potential for a large variety
of studies in Galactic astronomy. A summary list of today's
applications of wide binaries can include, in no particular order:

\begin{itemize}

\item probing for the processes and conditions of star formation as a
function of age and metallicity in star-forming regions (\cite[White
\& Ghez 2001]{whi01}), and as a function of environment (disk
vs. halo) during the assembly of the Galaxy (\cite[Chanam\'e \& Gould
2004]{cg});

\item to obtain new and accurate distances to faint low-mass stars in
the field (\cite[Gould \& Chanam\'e 2004]{hipp}; \cite[L{\'e}pine \&
Bongiorno 2006]{lep06}), allowing high-precision studies of stellar
properties near the bottom of the main sequence (\cite[Patten et
al. 2006]{pat06});

\item determination of the metallicities of field M-dwarfs
(\cite[Bonfils et al. 2005]{bon05}) and the study of the
age-metallicity relation in the Galactic disk (\cite[Monteiro et
al. 2006]{mon06});

\item tests of not well constrained stellar evolutionary processes
such as internal mixing and diffusion via the study of the surface
abundances of the members of wide binaries with twin components
(\cite[Mart\'{i}n et al. 2002]{mar02});

\item exploration of the age and chemical evolution of the stellar
halo and its underlying substructure via the study of halo wide
binaries with evolved components (Ivans, Chanam\'e, \& Gould, in
preparation);

\item measurement of the masses of white dwarfs and constraints on the
(dark) mass density of stellar remnants in the local Galactic disk
(\cite[Silvestri et al. 2001]{sil01});

\item to place severe constraints on the nature of halo dark matter
that complement the decades-long microlensing campaigns (\cite[Yoo,
Chanam\'e, \& Gould 2004]{yoo04}), and to explore the dynamical and
merger history of the Galaxy (\cite[Allen \& Poveda 2006]{all06});

\item studies of the initial-to-final mass relation of white dwarfs
via the determination of the mass of their main-sequence progenitors
(\cite[Catal\'an et al. 2006]{cat06}; \cite[Silvestri et
al. 2005]{sil05});

\item constraining mass loss during post main sequence stages, white
dwarf progenitor masses, and the density of matter returned to the
interstellar medium via the investigation of the evolution of orbital
separations of wide binaries harboring evolved components
(\cite[Johnston, Oswalt, \& Valls-Gabaud 2006]{joh06});

\item studying the impact of planetary systems on the evolution of
their host stars via the detailed surface abundances of stars in wide
binaries (\cite[Desidera et al. 2004]{des04}).

\end{itemize}

The main purpose of this review is therefore to highlight the large
potential inherent to {\it large samples} of wide
binaries\footnote{The discussion in the present contribution is
restricted to wide binaries in the field, i.e., does not include
systems in clusters and/or star forming-regions.} for the study of the
Galaxy. The emphasis on the words {\it large samples} is not
accidental: various of the applications mentioned above are not
necessarily the product of studies of wide binaries on a system by
system basis but, rather, statistically as a population. They have
only been possible thanks to the relatively large numbers of {\it
genuine} binaries that became available when astronomers changed their
strategies and started to look for them in a systematic way, making
big efforts to avoid possible selection biases. This change of
approach, in turn, only became possible with the construction and
exploitation of large photometric and astrometric databases such as
the NLTT, USNO, {\it Hipparcos}, 2MASS, etc. Therefore, the critical
role of {\it large surveys} in this revolution, not only in the field
of wide binaries but for Galactic astronomy in general, cannot be
emphasized enough, and this crucial point is stressed again in \S\,2
and throughout the present contribution.

In \S\,3 a brief report is presented of the status of the currently
available catalogs of wide binaries, using the case of the Revised
NLTT (rNLTT; \cite[Chanam\'e \& Gould 2004]{cg}) catalog to illustrate
the steps involved in their construction. A description of the ongoing
efforts and prospects of using the Sloan Digital Sky Survey (SDSS) to
build an even larger catalog of wide binaries is presented in
\S\,3.2. Finally, a few of the various applications of catalogs of
wide binaries to problems of Galactic astronomy are briefly outlined
in \S\,4.

\section{Surveys: the crucial starting point}
\label{sec:surveys}

Progress in astronomy over the last 25 or so years has been deeply
influenced by the steady increase in importance of the role of large
and systematic surveys of all types. Several of these even prompted
the deployment of instrumentation beyond Earth's atmosphere.  The
impact of these strategies is now seen across areas that range from
the cosmological (HST, COBE, 2dF, ...) to that of extrasolar planets
(SuperWasp, Kepler). In the Galactic context, the extraordinary
explosion of new science and discoveries that the SDSS (a survey
mainly designed to address cosmological questions) is allowing at this
very moment has its most immediate predecessors, both scientific as
well as technological, in surveys such as {\it Hipparcos} and 2MASS.
Today, the case for staying on the same track is evidenced by the
efforts (and funding!) put in upcoming surveys such as GAIA.

Research on wide binaries is no exception to this trend and progress
in this area is also intimately tied to the development and use of
large databases. Commonly referred to as common proper-motion systems,
a large number of wide binaries is a natural and relatively
straightforward product of astrometric campaigns and proper-motion
surveys.

Moreover, the combination of good photometry with good astrometry,
either coming from a single survey or from the combination of separate
ones, has allowed in the last few years the clean separation between
large numbers of stars unambiguously belonging to the disk and halo of
the Galaxy, thus opening a new avenue for studying Galactic structure
and evolution with high statistical significance. On the subject of
our interest here, the rNLTT catalog of disk and halo wide binaries,
assembled from the combination of surveys such as the NLTT, USNO-A,
{\it Hipparcos}, and 2MASS, constitutes an excellent example of this
synergy, and its capacity of telling whether a single star belongs to
the disk or the halo of the Galaxy constitutes a powerful tool for the
unambiguous identification of pairs of stars that, even though having
proper motions sharing the same magnitude and direction on the sky,
are not really bound to each other, and would contaminate a sample of
wide binaries selected only on common proper-motion criteria (see
\S\,3.1).

The big advantage of the strategy of surveys lies not only in the
greatly improved statistics but, most importantly, in the degree of
completeness of the resulting databases, which, with well understood
selection effects, if any, permits a relatively unbiased view of
whatever the particular survey aims at. It is this crucial
characteristic that, in the case of wide binaries, makes possible two
of their most interesting applications to astronomy, as are their use
as probes of the nature of dark matter and of the processes and
conditions of star formation (see \S\,4).

The approach of large surveys has proven to be of great power and is
likely to stay as one of the favored ways to advance in
astrophysics. People interested in wide binaries and their
applications must therefore consider carefully what can be done with
the already existing databases as well as prepare in advance for an
efficient use and exploitation of the surveys to come. Joint
Discussion 13 on this very same IAU General Assembly, entitled {\it
Exploiting Large Surveys for Galactic Astronomy}, dealt with this
subject in detail, and readers are encouraged to consult those
proceedings (see also their website at
http://clavius.as.arizona.edu/vo/jd13/).

\section{Catalogs of field wide binaries}
\label{sec:catalogs}

As noted before, being the characteristic fingerprint of wide binaries
the almost identical proper-motion vectors of their two stars (up to
differences due to measurement uncertainty and relative orbital motion
only), they are intimately associated to proper-motion surveys. The
rNLTT wide binaries, extracted from the Luyten survey of high
proper-motion stars (\cite[Luyten 1979, 1980]{luy}), constitute likely
the most homogeneous and complete catalog of such objects available
today, although not the only one.

Before the rNLTT, the most successful efforts to assemble large
numbers of wide binaries were those of Poveda et al. (1994) and Allen
et al. (2000), who based their searches on databases such as Gliese's
Catalog of Nearby Stars and the NLTT itself, though not exploiting the
latter in its full capacity. See the contributions by A. Poveda and
C. Allen on these proceedings for details on their samples and some of
their uses in astronomy. The efforts by \cite{all00} and \cite{zap04}
concentrated on the search for wide companions to metal-poor stars
selected from databases such as the Carney-Latham surveys. A few
earlier attempts to identify smaller samples of field wide binaries
are reported in the introductory section of \cite{cg}.

Other proper-motion catalogs available today that hold potential for a
search of wide binaries include UCAC (\cite[Rafferty et
al. 2001]{raf01}), SuperCOSMOS (\cite[Hambly et al. 2001]{ham01}),
LSPM (\cite[L\'epine \& Shara 2005]{lspm}), and SDSS$\,\cap\,$USNO-B
(\cite[Gould \& Kollmeier 2004]{gou04}; \cite[Munn et
al. 2004]{mun04}). Although a few new common proper-motion pairs of
intrinsic interest have been identified from these databases
(\cite[Scholz et al. 2002, 2005]{sch02}; \cite[Seifahrt et
al. 2005]{sei05}; \cite[Monteiro et al. 2006]{mon06}), only one report
of a systematic search for wide binaries using these catalogs exists,
which appeared the very same day this contribution was to be
submitted. In a manner analogous to the search of \cite[Gould \&
Chanam\'e (2004)]{hipp} in the rNLTT, \cite[L\'epine \& Bongiorno
(2006)]{lep06} searched in the LSPM catalog for faint common
proper-motion companions of {\it Hipparcos} stars, uncovering 521
systems, of which 130 are new (see \S\,4).

The optimal construction of a catalog of field wide binaries is
outlined in \S\,3.1, using the case of the rNLTT as an illustrative
example. It will be seen how, with appropriate photometric and
astrometric data, it is possible to not only identify genuine field
wide binaries, but also to determine with high confidence, on a pair
by pair basis, their membership in either the disk or the halo
populations of the Galaxy. Disentangling between pairs belonging to
the Galactic thin and thick disks, however, would require more
information, such as the full three-dimensional velocities and
metallicities of the stars. Finally, in \S\,3.2, I describe an ongoing
search for wide binaries in SDSS$\,\cap\,$USNO-B.

\subsection{Building a catalog: the rNLTT wide binaries}

The NLTT proper-motion survey was available long before the \cite{cg}
work, and Luyten himself identified and recorded a substantial
fraction of the NLTT wide binaries in his Luyten Double-Star Catalog
(LDS, \cite[Luyten 1940-87]{lds}). However, two factors worked against
the construction of a {\it complete} sample of wide binaries from the
original NLTT. First, the crude photographic colors did not allow the
construction of a reliable reduced proper motion (RPM) diagram, thus
preventing the identification of chance alignments (i.e., pairs of
unrelated stars whose proper-motion vectors are aligned on the sky
just by chance) via the non-consistent colors of the stars in such
pairs (more below). Second, the large proper-motion errors in the
original NLTT did not allow Luyten to identify a fair number of
genuine binaries at large angular separations (the {\it non-Luyten}
binaries of \cite[Chanam\'e \& Gould 2004]{cg}) because it was
impossible to separate them from the numerous unrelated optical pairs
at those separations.

\begin{figure}
\hspace{0cm}\includegraphics[height=2.8in,width=2.8in,angle=0]{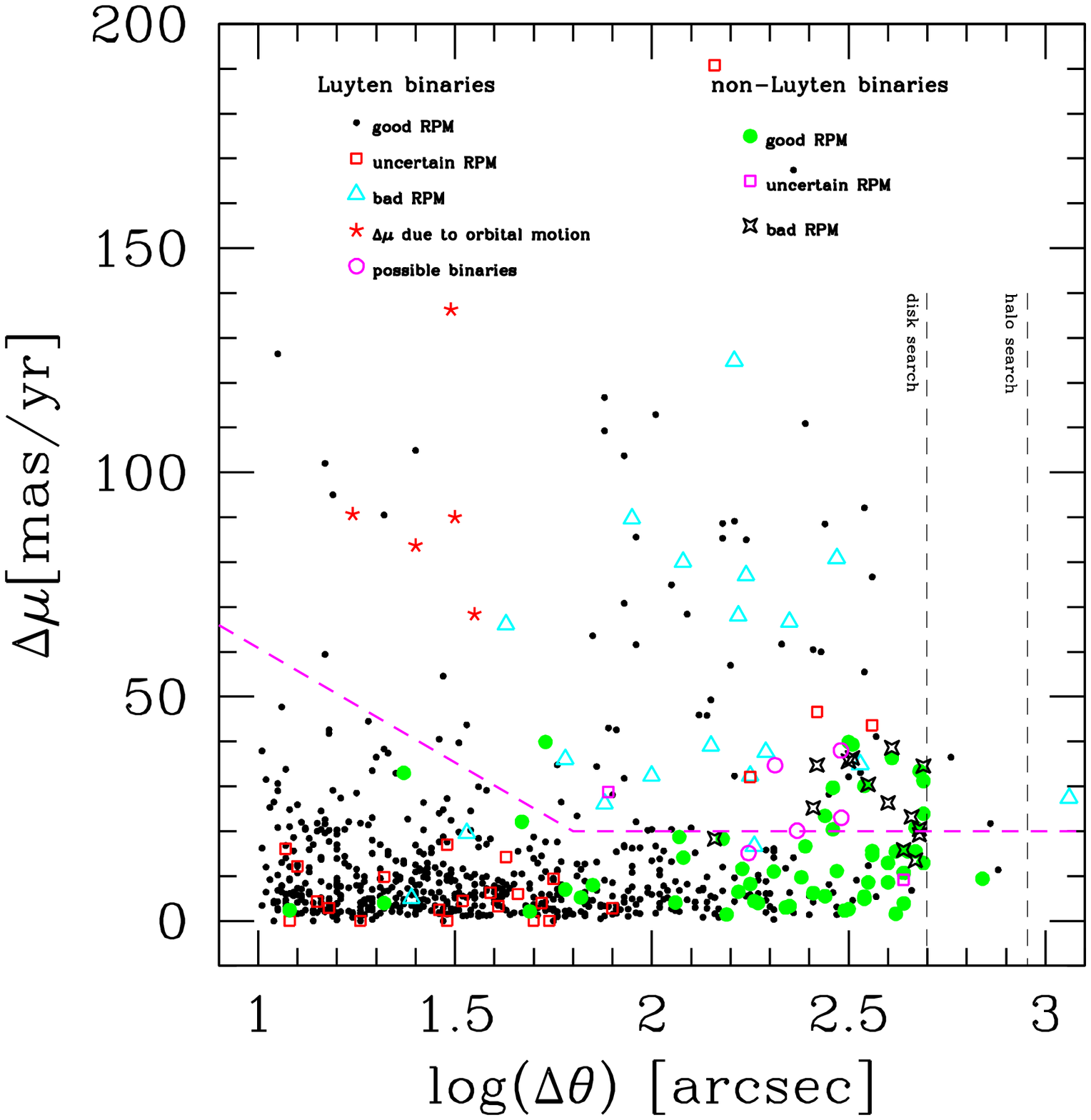}
\end{figure}

\begin{figure}
\vspace{-7.5cm}
\hspace{6.7cm}\includegraphics[height=2.85in,width=2.85in,angle=0]{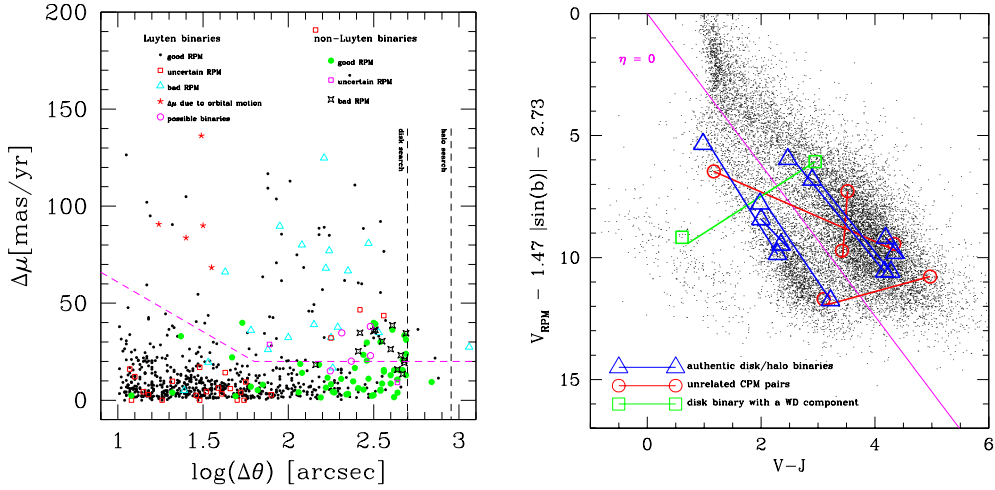}
\caption{(a; left) Initial selection of candidate binaries in rNLTT:
vector proper motion difference vs. angular separation for systems
with $\Delta\theta > 10\arcsec$. Pairs below the dashed line were
automatically accepted as candidate binaries for their subsequent
classification. The dashed vertical lines indicate the upper limits of
the homogeneous search for disk and halo binaries. \break (b; right)
Reduced proper motion diagram of rNLTT stars, clearly separating the
disk, halo, and white dwarf tracks. A few examples of how this diagram
is used to test and classify binaries are shown (see text).  }
\label{fig:search}
\end{figure}

Only with the construction of the Revised NLTT (\cite[Gould \& Salim
2003]{bright}; \cite[Salim \& Gould 2003]{faint}) could these problems
be overcome. An upgraded version of the original NLTT, the rNLTT
catalog provides optical and (CCD) infrared photometry for the
majority of the NLTT stars by matching them to the USNO-A and 2MASS
surveys. Thanks to the better colors and large color baseline, this
proved sufficient to permit a clean separation between disk and halo
stars using a RPM diagram (\cite[Salim \& Gould
2002]{rpm}). Furthermore, the better temporal baseline of the rNLTT
relative to the NLTT translated into a substantial improvement in the
accuracy of the proper-motion measurements, which allowed
\cite[Chanam\'e \& Gould (2004)]{cg} to perform a uniform and complete
search of wide common proper-motion pairs up to angular separations
much larger than what Luyten was able to do. It is worth noticing,
here again, the fundamental importance of large surveys and the
potential inherent in their combination and adequate exploitation.

Still, with the main shortcomings of the original NLTT overcome with
the publication of the rNLTT, it was by no means straightforward to
dig out a {\it uniformly selected} catalog of wide binaries from
it. In order to produce a sample of wide binaries as free from
selection effects as possible, a number of complications had to be
dealt with, such as blending, saturation, missing independent
proper-motion measurements, and others. See \cite{cg} for a detailed
account of the solutions to these problems. Here only a brief outline
of the selection and classification of the rNLTT wide binaries is
given.

The first step is to select all pairs of stars that seem to be moving
together on the sky, i.e., all pairs with similar proper-motion
vectors. This was done by plotting the magnitude of the vector proper
motion difference (denoted by $\Delta{\rm\bf\mu} =
|\Delta{\rm\mbox{\boldmath$\mu$}}|$) of the components of these pairs
against their angular separation ($\Delta\theta$), as shown in Figure
1a. Pairs with separations smaller than $\Delta\theta = 10$\arcsec as
measured by Luyten were not subjected to this selection because of the
small chance of two unassociated stars lying so close in both position
and velocity space. Thus the plot in Figure 1a begins at $\Delta\theta
= 10$\arcsec and for pairs with smaller separations all efforts were
focused on their classification (which, based on the consistency of
their colors, constitutes another test for their reality).

For the rest of common proper-motion systems wider than 10\arcsec,
good candidates were considered to be those falling below the dashed
line in Figure 1a, set at wide separations at $\Delta{\rm\bf\mu} =
20\masyr$. Note that all the NLTT stars are moving faster than
$180\masyr$, so the above maximum-$\Delta\mu$ requirement is already a
strong test, and pairs that satisfy it have already a high likelihood
of being real binaries. The $20\masyr$ cutoff was somewhat relaxed at
closer separations because the chance of contamination by unassociated
pairs is lower and also because some real pairs at close separations
can have significant orbital motions (pairs indicated by red asterisks
in Fig. 1a).

This search for rNLTT wide binaries was systematically done up to
separations of $\Delta\theta = 500$\arcsec for disk binaries and
$\Delta\theta = 900$\arcsec for halo binaries, thus avoiding any
possible selection effects as a function of angular separation. In the
same spirit, the exclusion from the analysis of pairs outside the
allowed region of the $\Delta\theta - \Delta\mu$ plane does not mean
that all these systems should be regarded as false binaries. Rather,
it simply reflects the extreme care put in being as complete and free
from selection effects as possible, but at the same time rigorous in
not being contaminated by unrelated optical pairs. It is only thanks
to this characteristic that the rNLTT wide binaries (or any other
catalog similarly constructed) can be used as probes of star formation
as well as of the nature of dark matter in the Galaxy.

The second test for the reality of the (up to this point) candidate
binaries is related with their classification as either disk or halo
pairs. This is done via examination of each pair in a RPM diagram, and
all the possibilities are illustrated in Figure 1b, where the clear
separation between the tracks of disk and halo stars (at both sides of
the line labeled $\eta = 0$) is evident. When classifying the
binaries, one expects not only that both components belong to the same
population, but also that their positions on the RPM diagram are
consistent with both stars having the same age and metallicity.
Therefore, for real binaries (blue triangles in Fig. 1b) one expects
that the line connecting their positions on the RPM diagram should be
approximately parallel, within measurement errors, to the
corresponding disk or halo track. Candidate pairs composed of one disk
and one halo member are rejected as chance alignments, and so are
pairs composed of two disk or two halo stars if the line connecting
them is inconsistent with being parallel to the respective sequence
(red circles in Fig. 1b). The only cases permitted not to follow this
parallel rule are those involving a white dwarf companion (as the
green pair of squares in Fig. 1b).

The above two-step procedure yielded 999 genuine wide binaries: 883
pairs belonging to the Galactic disk and 116 belonging to the Galactic
halo. Among the disk binaries, 82 pairs contained a white dwarf, while
no halo stars with clear white dwarf companions were found.

\subsection{Wide binaries from the SDSS$\,\cap\,$USNO-B}

The Sloan Digital Sky Survey, providing superb optical CCD photometry
for almost a quarter of the sky, is already revolutionizing our
understanding of the Galaxy on several fronts. Although a single-epoch
survey, it can be combined with older photometric and/or proper-motion
surveys in order to produce the two epochs necessary to compute new
proper motions for its stars. \cite[Gould \& Kollmeier (2004)]{juna}
and \cite[Munn et al. (2004)]{mun04} already accomplished this for the
SDSS First Data Release by cross-correlating it with the USNO-B
catalog, producing nearly 400,000 stars down to $r' < 20$, moving
faster than $20\masyr$, and with proper-motion errors of about
$4\masyr$.

Given the depth and sky coverage of SDSS, there is little doubt that
it will provide a significantly larger number of binaries than the
rNLTT, and efforts toward this goal are already underway.  The
challenge, however, lies in being able to distinguish the truly bound
systems from among the large number of spurious pairs inevitably
present in such a large collection of stars with very similar proper
motions.  For the nearby Luyten stars, with proper motions
$|\rm\mbox{\boldmath$\mu$}| > 180\masyr$, just the combination of
photometry and astrometry was enough to do the selection.  But with
the more distant SDSS stars, moving at $20-40\masyr$, the measured
$\Delta\vec{\mu}$ alone does not reject all the apparent binaries,
even though the photometry is excellent.  Therefore, it is necessary
to obtain radial velocities of the components of the common
proper-motion candidates that pass the first tests (consistent
photometry and proper motions) to clean the initial sample and select
the real binaries. Binary candidates have been selected and the radial
velocities of the brightest pairs have already been measured using the
SMARTS 1.5m telescope at CTIO (Chanam\'e, van der Marel, van de Voort,
\& Gould, in preparation). The large majority of targets, however,
with magnitudes in the range $16 < r < 20$, require larger
apertures. Our plans to complete these observations using telescopes
at the MDM and Kitt Peak Observatories are already underway.

\section{Impact of wide binary catalogs on Galactic astronomy}
\label{sec:applications}

\begin{figure}
\hspace{0cm}\includegraphics[height=2.1in,width=2.5in,angle=0]{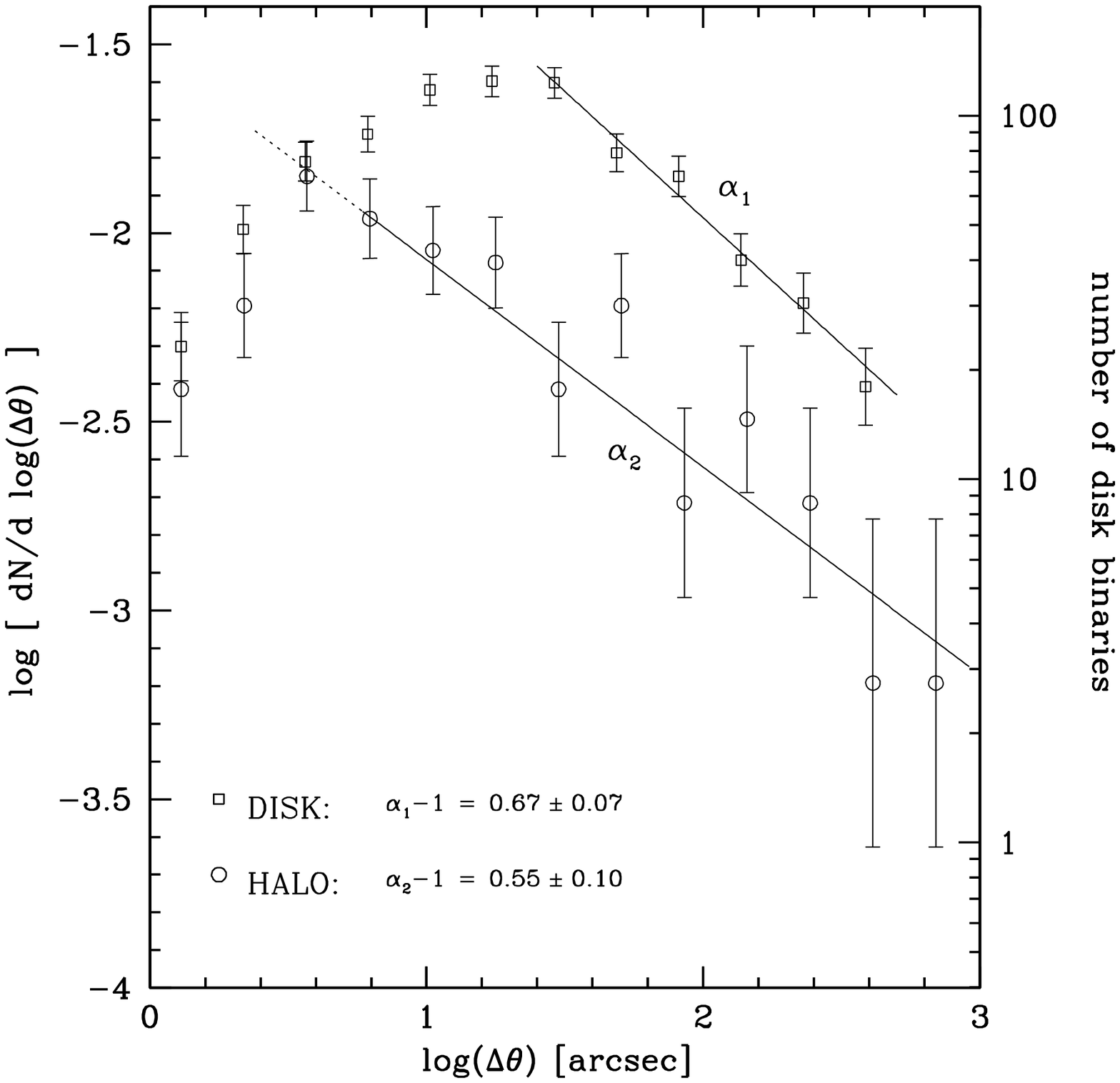}
\end{figure}

\begin{figure}
\vspace{-5.15cm}
\hspace{6.5cm}\includegraphics[height=1.8in,width=2.8in,angle=0]{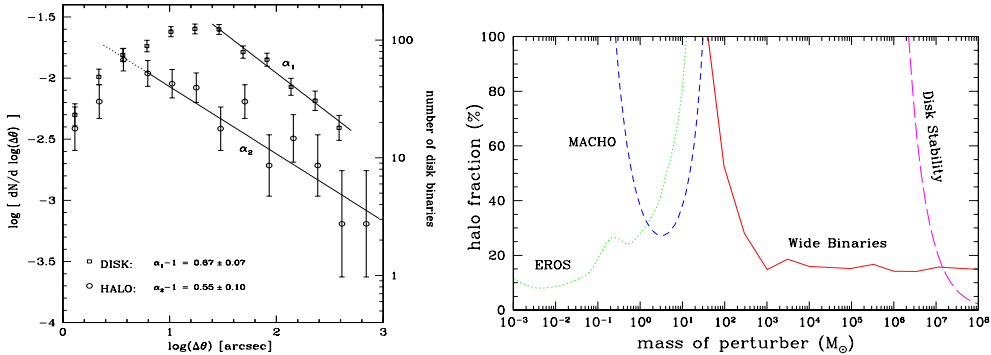}
\caption{(a; left) Distribution of angular separations for the rNLTT
disk and halo wide binaries (\cite[Chanam\'e \& Gould 2004]{cg}). Note
that disk and halo binaries show basically the same distributions, and
that the data for both populations of binaries are consistent with a
single power-law out to the widest separations, with no hint of a
break or cutoff. (b; right) Exclusion contours for MACHO dark matter
at the $2\sigma$ level of the microlensing experiments and the rNLTT
halo wide binaries (\cite[Yoo, Chanam\'e, \& Gould 2004]{yoo04}).  }
\label{fig:machos}
\end{figure}

Here, I only point out a few of the various applications of wide
binaries to Galactic astronomy that are outlined in \S\,1. See
T. Oswalt's contribution for the several uses of wide binaries
involving white dwarfs, and those of C. Allen and A. Poveda for ways
of exploiting their potential as probes of dynamical processes in the
Galactic disk.

One of the most useful characteristics of wide binaries is that both
components share properties such as age, metallicity, distance, etc.,
and, therefore, when any of these is measured for one of them,
typically the primary, it automatically can be assigned to the fainter
secondary.  This has been exploited to obtain parallaxes to faint
low-mass stars in the field by \cite{hipp} and \cite{lep06}, who
searched for wide companions of {\it Hipparcos} stars in the rNLTT and
LSPM catalogs, respectively. Other applications listed in \S\,1 apply
this same principle to get ages, metallicities, etc.

Nevertheless, two of the most exciting applications of wide binaries
are the result of statistical studies for which not only the large
size but, perhaps most importantly, the completeness of any sample of
these objects is critical. In particular, their distribution of
(physical or angular) separations can be used to gain insights into
the processes of star formation, and also to place strong limits to
the mass and density of compact dark matter (MACHOs) in the Galactic
halo.

Using the rNLTT, \cite{cg} were able to perform the first comparison
of the distributions of separations of large and clean samples of disk
and halo wide binaries, as shown in Figure 2a.  Somewhat surprisingly,
they found both distributions to be the same within the uncertainties,
despite the radically different Galactic environments to which they
belong today. This provides a unique insight on the environmental
conditions in which star formation proceeded in the early Galaxy, and
any successful model of the formation of the Galaxy should be able to
reproduce this finding.

Finally, the very wide binaries ($a \gtrsim 0.1$ pc) are so weakly
bound that they can be significantly disturbed, even disrupted, by
gravitational interactions as they orbit the Galaxy, including those
with giant molecular clouds, large-scale tides, passing stars, as well
as massive dark objects (MACHOs). If such encounters were frequent
during the lifetime of any given binary, then one would expect the
widest systems to be relatively depleted. Early attempts to learn
about disk dark matter in the solar neighborhood via wide binaries
(\cite[Bahcall, Hut \& Tremaine 1985]{bht}) did not achieve useful
conclusions due to the small number of wide binaries available then,
as well as the intrinsic complexity of the disk potential
(\cite[Weinberg, Shapiro \& Wasserman 1987]{martin}).

The situation in the Galactic halo was easier to model, but an
adequate sample of halo wide binaries did not exist until the rNLTT
binaries were published. Only then, using detailed Monte Carlo
simulations of the evolution of a population of wide binaries immersed
in a halo filled with compact objects with various combinations of
mass and density, \cite[Yoo, Chanam\'e, \& Gould (2004)]{yoo04} were
able to exclude a Milky Way halo mostly composed of MACHOs with masses
$M > 43$\msun at the standard local halo density and at $95\%$
confidence level, as shown in Figure 2b. In this way, halo wide
binaries were shown to complement the results of the microlensing
campaigns, mostly sensitive to lower masses, and provided us with an
alternative experiment for addressing the same key question.

\begin{acknowledgments}

I wish to thank Bill Hartkopf for inviting me to this Symposium, and
Roeland van der Marel for his interest and support of my work on wide
binaries. Thanks as well to Andy Gould and Inese Ivans for a critical
reading of this manuscript, and to Freeke van de Voort for her help
and enthusiasm in digging into noisy spectra in the search for new
binaries in SDSS.

\end{acknowledgments}



\begin{thebibliography}{}

\bibitem[Allen \& Poveda (2006)]{all06} Allen, C., \& Poveda, A.\ 2006, IAU Symposium, 240 (these proceedings)
\bibitem[Allen et al. (2000)]{all00} Allen, C., Poveda, A., \& Herrera, M.\ 2000, \textit{A\&A}, 356, 529
\bibitem[Bahcall, Hut \& Tremaine (1985)]{bht} Bahcall, J. N., Hut, P., \& Tremaine, S. 1985, \textit{ApJ}, 290, 15
\bibitem[Bonanos et al. (2006)]{bon06} Bonanos, A.~Z., et al.\ 2006, \textit{Ap\&SS}, 79
\bibitem[Bonfils et al. (2005)]{bon05} Bonfils, X., et al.\ 2005, \textit{A\&A}, 442, 635
\bibitem[Catal\'an et al. (2006)]{cat06} Catal\'an, S., et al.\ 2006, IAU Symposium, 240 (these proceedings)
\bibitem[Chanam\'e \& Gould (2004)]{cg} Chanam\'e, J. \& Gould, A.\ 2004, \textit{ApJ}, 601, 289
\bibitem[Desidera et al. (2004)]{des04} Desidera, S., et al.\ 2004, \textit{A\&A}, 420, 683
\bibitem[Drake, Cook, \& Keller (2004)]{drake04} Drake, A.J., Cook, K.H., \& Keller, S.C.\ 2004, \textit{ApJ}, 607, L29
\bibitem[Gould \& Chanam\'e (2004)]{hipp} Gould, A., \& Chanam\'e, J.\ 2004, \textit{ApJS}, 150, 455
\bibitem[Gould \& Kollmeier (2004)]{juna} Gould, A., \& Kollmeier, J.\ 2004, \textit{ApJS}, 152, 103
\bibitem[Gould \& Salim (2003)]{bright} Gould, A., \& Salim, S.\ 2003, \textit{ApJ}, 582, 1001
\bibitem[Hambly et al. (2001)]{ham01} Hambly, N.~C., et al.\ 2001, \textit{MNRAS}, 326, 1315
\bibitem[Jiang et al. (2004)]{guangfei} Jiang, G., et al.\ 2004, \textit{ApJ}, 617, 1307
\bibitem[Johnston et al. (2006)]{joh06} Johnston, K., Oswalt, T., \& Valls-Gabaud, D.\ 2006, IAU Symposium, 240 (these proceedings)
\bibitem[L{\'e}pine \& Shara(2005)]{lspm} L{\'e}pine, S., \& Shara, M.~M.\ 2005, \textit{AJ}, 129, 1483
\bibitem[L{\'e}pine \& Bongiorno (2006)]{lep06} L{\'e}pine, S., \& Bongiorno, B.\ 2006, astro-ph/0610605
\bibitem[Luyten (1979, 1980)]{luy} Luyten, W.\ J.\ 1979, 1980, New Luyten Catalog of Stars with Proper Motions Larger than Two Tenths of an Arcsecond (Minneapolis: University of Minnesota Press)
\bibitem[Luyten (1940-87)]{lds}Luyten, W.J.\ 1940-87, The LDS Catalogue: Double Stars with Common Proper Motion (Minneapolis: University of Minnesota Press)
\bibitem[Mart{\'{\i}}n et al. (2002)]{mar02} Mart{\'{\i}}n, E.L.,\ 2002, \textit{ApJ}, 579, 437
\bibitem[Monteiro et al. (2006)]{mon06} Monteiro, H., Jao, W.-C., Henry, T., Subasavage, J., \& Beaulieu, T.\ 2006, \textit{ApJ}, 638, 446
\bibitem[Munn et al. (2004)]{mun04} Munn, J.~A., et al.\ 2004, \textit{AJ}, 127, 3034
\bibitem[Paczy\'nski (1997)]{pac97} Paczy\'nski, B.\ 1997, The Extragalactic Distance Scale, 273 (astro-ph/9608094)
\bibitem[Pagel (1997)]{pag97} Pagel, B.\ 1997, Nucleosynthesis and Chemical Evolution of Galaxies (Cambridge Univ Press)
\bibitem[Patten et al. (2006)]{pat06} Patten, B.M., et al.\ 2006, astro-ph/0606432
\bibitem[Poveda et al. (1994)]{pov94} Poveda, A., et al.\ 1994, Revista Mexicana de Astronom\' ia y Astrof\' isica, 28, 43
\bibitem[Rafferty et al. (2001)]{raf01} Rafferty, T.~J., et al.\ 2001, ASP Conf.~232: The New Era of Wide Field Astronomy, 232, 308
\bibitem[Ribas et al. (2005)]{rib05} Ribas, I., et al.\ 2005, \textit{ApJ}, 635, L37
\bibitem[Salim \& Gould (2002)]{rpm} Salim, S., \& Gould, A.\ 2002, \textit{ApJ}, 575, L83
\bibitem[Salim \& Gould (2003)]{faint} Salim, S.~\& Gould, A.\ 2003, \textit{ApJ}, 582, 1011
\bibitem[Scholz et al. (2002)]{sch02} Scholz, R.-D., et al.\ 2002, \textit{ApJ}, 565, 539
\bibitem[Scholz et al. (2005)]{sch05} Scholz, R.-D., et al.\ 2005, \textit{A\&A}, 430, L49
\bibitem[Seifahrt et al.(2005)]{2005A&A...440..967S} Seifahrt, A., et al.\ 2005, \textit{A\&A}, 440, 967
\bibitem[Silvestri et al. (2001)]{sil01} Silvestri, N., et al.\ 2001, \textit{AJ}, 121, 503
\bibitem[Silvestri et al. (2005)]{sil05} Silvestri, N., Hawley, S.~L., \& Oswalt, T.~D.\ 2005, \textit{AJ}, 129, 2428
\bibitem[Weinberg, Shapiro \& Wasserman (1987)]{martin} Weinberg, M.D., Shapiro, S.L., \& Wasserman, I. 1987, \textit{ApJ}, 312, 367
\bibitem[White \& Ghez (2001)]{whi01} White, R.~J., \& Ghez, A.~M.\ 2001, \textit{ApJ}, 556, 265
\bibitem[Yoo, Chanam\'e, \& Gould (2004)]{yoo04} Yoo, J., Chanam\'e, J., \& Gould, A.\ 2004, \textit{ApJ}, 601, 311
\bibitem[Zapatero Osorio \& Mart{\'{\i}}n (2004)]{zap04} Zapatero Osorio, M.~R., \& Mart{\'{\i}}n, E.~L.\ 2004, \textit{A\&A}, 419, 167


\end{thebibliography}
\end{document}